# Modeling of High-Gradient Vacuum Breakdown

A. Ainabayev[1], B. Berdenova[1], J. Norem [2, a)], Z. Insepov [1, 3, b)]

[1] Nazarbayev University Research and Innovation System, Astana, Kazakhstan
[2] Nano Synergy, Inc., IL USA
[3] Purdue University, West Lafayette, IN USA

Corresponding authors: [a)] norem.jim@gmail.com, [b)] zinsepov@purdue.edu

**Abstract.** Comsol finite element technique was applied to study heating and cooling of a microspot on the cathode surface. The reasons why there seems to be no common model for vacuum arcs, in spite of the importance of this field and the level of effort expended over more than one hundred years, were explored.

## WHY IS PROGRESS SO SLOW?

While vacuum arcs were first identified around 1900, by Michelson and Millikan (before they even had a vacuum pump), the first reasonable explanation of the phenomenon was proposed by Lord Kelvin in 1904, and the field has been continuously active since then, there has been no general agreement on the nature of these arcs that has developed in the past 110+ years. There are a number of both experimental and theoretical problems that seem to have retarded progress.

Experimentally, the arcs develop very fast and rapidly obliterate the surface defects that presumably help trigger the arcs, and the rapid arc development, over many orders of magnitude in many parameters, presents a problem with diagnostics, which tend to be useful over a comparatively small parameter range. In addition, arcing tends to occur unpredictably, both in location and time, frequently in locations (Tokamaks, accelerators, etc. where access is poor) presenting further diagnostic problems. Finally, there is a wide variety of arcs and related phenomena that may or may not be related to basic arc mechanisms, for example in micrometeorite impacts with satellites, Tokamak plasma/limiter interactions, the rf accelerator limits we are familiar with, laser ablation, and other applications. Even within a specific type of arc, the experimental parameters,(stored energy, pulse length, geometry, materials) would be expected to vary from one experiment to another. An additional problem, however is the complexity and variety of the mechanisms involved, which seem to operate over parameter ranges of many orders of magnitude and a diversity of environments.

We have found that in order to model arcs, it seems necessary to incorporate many highly specific and somewhat incompatible calculations and calculation methods, for example Particle In Cell (PIC) codes to study the initial stages of the arc, and Molecular Dynamics (MD) to look at the properties of the fully developed arc. Surface effects are also an integral part of this problem, however surface structure and surface issues are not generally considered a part of the plasma physics of arcs. Computational methods assume different boundary conditions, different timescales and different internal dynamics, and it is not obvious that they are simply compatible. We find that the concept of the unipolar arcs seems to have wide applicability, however the application of these methods to his problem is not straightforward and the literature on these arcs is not well developed or unambiguous. Application of MD methods is a slow process; our paper looking at the application of MD to the properties of the dense (non-ideal) plasma sheath seems to be the first that makes definite predictions of the surface physics at the edge of these non-ideal plasmas, however it is not clear how to experimentally verify these predictions.

The arc process consists of a number of elements. During the arc the energy in the electric field is transferred to the heat in the wall. This paper summarizes calculations of surface heating under a number of models of arcs and beam optics.

## MODELING PHYSICAL MECHANISMS IN ARCING

In accelerator RF systems arcs have dimensions measured in mm, survive for times on the order of 10 ns to 100 $\mu$s, and involve energies measured in Joules. The mechanisms operating within arcs operate on much different spatial, time and energy scales, and the parameters of the individual mechanisms determine the type of numerical analysis that can be done.

Vacuum arcing seems to be dominated by a number of quite different mechanisms, representing different fields of study. For example, the arc seems to be described by plasma physics, however the initial breakdown stages are a complicated mixture of field emission (quantum mechanics) electrostatics, atomic physics, mechanical properties, and fracture mechanics. The surface damage that seems to determine the location of mechanical failure seems to be a result of hydrodynamics and thermodynamics. External factors also control many aspects of the arc, in particular its initiation and duration. Thus, we do not expect to see both simplicity and precision in initial modeling results.

There are a number of general methods involved in arc studies and we can explore the properties and limitations of each of them. While they do not present insolvable computational problems, seem to require a set of basic assumptions to unite an array of analyses of specific mechanisms.

## Particle in Cell

Classical plasma physics represents the core of the arcing problem. Particle in cell calculations are a well understood method of modeling plasma physics for linear systems with two body collisions, finite numbers of particles, timescales on the order of a few nanoseconds and geometrical dimensions on the order of $10^{-5}$ m. We have used PIC codes to model the first few nanoseconds of the arc development, when the arc satisfies the constraints up to the limit where we do not trust simple expressions like the derivation of the Debye length, $\lambda$.

On the other hand it is not always clear how precise PIC code predictions are when the densities become large enough so that the basic assumptions of classical plasma physics no longer apply, in this case because the density of the plasma is too large to assume that the energy of the system is entirely kinetic, or the collisions are entirely between two particles.

## Molecular Dynamics

We have used MD to model the mechanical failure of asperities that trigger breakdown, the self-sputtering off of solid and liquid surfaces that ultimately fuel the arc, and the properties of the sheath for dense plasmas, where the assumptions of two body interactions no longer apply. Computational limitations imply timesteps for these calculations on the order of $10^{-18}$ - $10^{-17}$ s and equilibrium times of $10^{-13}$ s, over geometrical dimensions of $10^{-9}$ m, many orders of magnitude smaller than the dimensions of PIC calculations. On the other hand, MD calculations can be used when the densities are high enough so that the total electrostatic energy of the system is comparable to the kinetic energy of the particles (nonideality coefficient $\Theta =$ electrostatic/kinetic energy ~0.5). MD calculations are not, in principle, compatible with PIC calculations, thus we use them to define boundary conditions and evaluate sputtering coefficients rather than to describe the evolution of the system as a whole

## Various mesoscale thermodynamics and other methods

A number of other mechanisms also contribute to arcing with their own parameter ranges and variables. Here we present our recent simulation results obtained by finite element Comsol Multiphysics software package [1].

Vacuum breakdown is a phenomenon of formation of charge carriers, usually electrons, in the vacuum space between two electrodes, caused by applying a high voltage that is above certain critical limit.

Vacuum breakdown is a complex, and not yet fully modeled. However, there are a few arguments that can explain various stages of the breakdown.

According to field emission theory, a high electric gradient can reduce the potential barrier at the metal surface that is keeping electrons inside the metal, and this allow some high kinetic energy electrons leave the surface and accelerate in the vacuum space in an electric field. These electrons having significant acceleration bombard the anode surface and can produce xrays.

Primary electrons then can ionize neutral atoms and thus form plasma. At certain concentration of plasma, energetic electrons can trigger the process impact ionization leading to an electron avalanche.

As a result, the concentration of plasma increases significantly and the ions start to drift in the direction of the cathode. Energetic ions further sputter the metal surface, thus further increasing the plasma concentration. At certain concentration of plasma, this leads to unipolar arc generation and discharge of the energy in the electric field. The energy in the electric field is then transferred into the wall, locally, in the form of heat.

In this paper, the dynamics of microscopic heating and cooling of a cupper cathode was studied by using a COMSOL Multiphysics finite element technique and the characteristic times of the processes were calculated.

Fig. 1 shows three geometries:

1) Surface is heated by a large area plasma, thus the heating and cooling processes occur in a 1-dimensional coordinate system (Fig. 1a).

2) Surface is heated by a collimated electron beam as would occur if the shorting current was focused by a magnetic field, cooling is simulation in two dimensions (Fig. 1b)

3) Surface is heated by a small localized plasma, cooling is simulated to occur in three dimensions. (Fig. 1c).

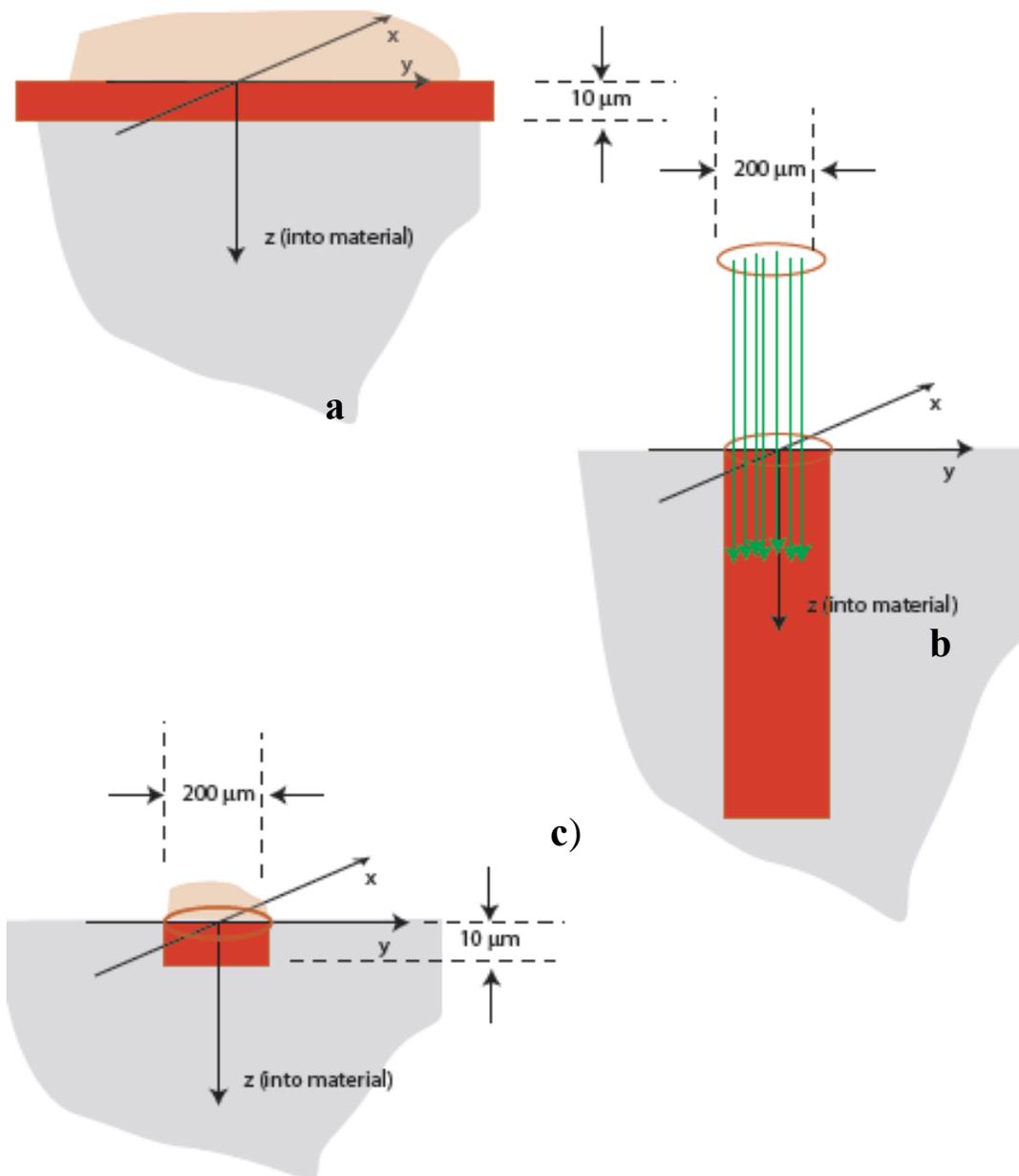

Fig. 1. Different geometries of heating and cooling simulated by Comsol Multiphysics finite element method.

According to the physical experimental conditions, heating was simulated until the copper melting temperature of 1350K, the system was held at this temperature for about 100 ns (fig. 2a ) with further thermal expansion (fig. 2c ) and cooling was simulated within the following 100 ns (fig. 2b ). If a crated was simulated (fig. 2d), a pressure inside the crated was established to be of 100 Pa. As a result, the following simulation instances were obtained in Comsol Multiphysics (fig 2).

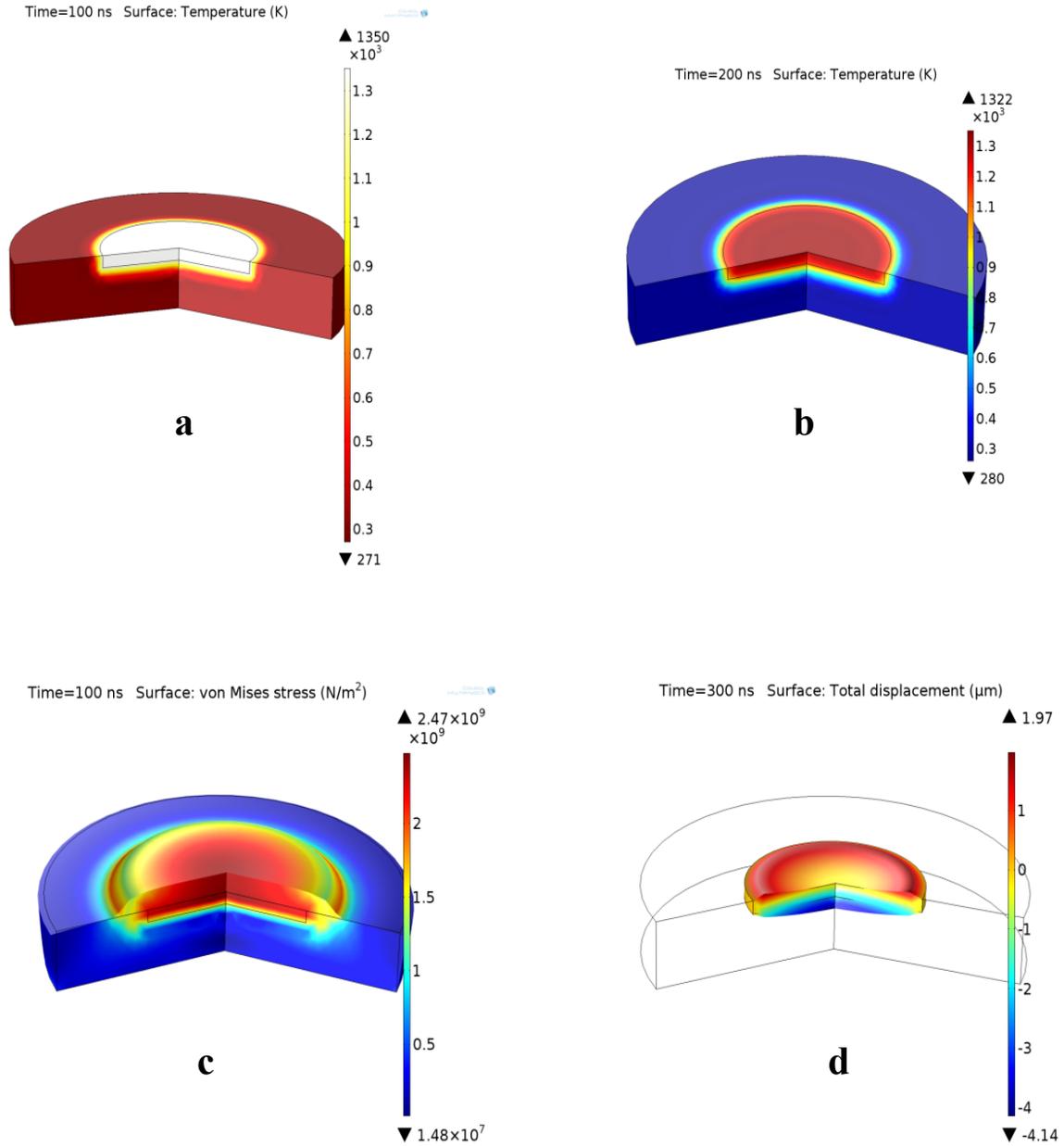

Fig. 2. Heating and cooling simulated by COMSOL Multiphysics package (for 3d geometry): a) heating process of cupper; b) cooling process of cupper; c) thermal expansion of cupper; d) crater formation; e) cooling of crater.

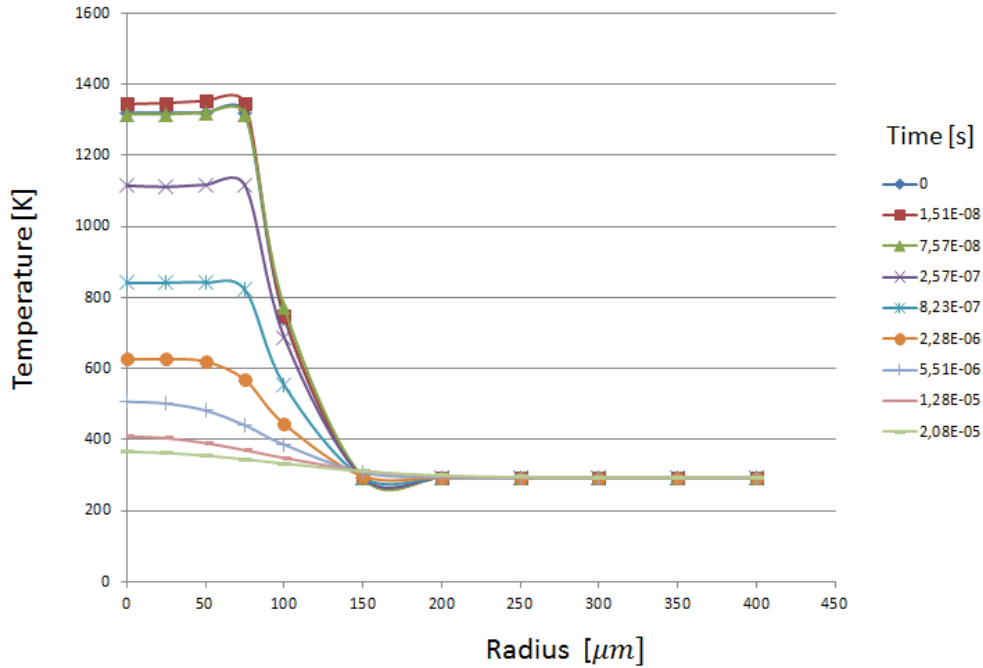

Fig. 3 Result of simulation of cooling in X and Y directions for 3d geometry.

## NEED FOR REALISTIC MODELING

Since the problem of arcs has been studied for over 100 years, a simple explanation of the physics of this phenomenon seems overdue. Beyond this, arcing and gradient limits in general are a significant constraint on many aspects of technological progress and ignorance of the primary mechanisms may be associated with significant costs. We find that many of the models in print do not seem to agree with current data, and can be incompatible with basic assumptions and modern modeling.

There are a variety of phenomena that seem to require simple explanations, such as the "chicken track" damage left by Tokamak arcs, the voltage spikes produced during micrometeorite impacts on satellites and the sensitivity of RF gradient limits to strong magnetic fields, and while it might seem desirable to have a complete numerical model of arcing, the complexity of this problem seems to preclude this. On the other hand, a complete understanding of the individual mechanisms involved, with numerical models that produce reasonable estimates of the critical variables i, seems to be something that can realistically be done. This goal seems obtainable.

## CONCLUSIONS

Because of the dynamic ranges involved, complete computational methods used to model arcing from a knowledge of the basic mechanisms are inherently problematic, because of widely different timescales, spatial volumes and plasma parameters. Thus modeling of arcing cannot, at present, be done using a single computational model, and must essentially be a parameter list obtained from a variety of models each applying to one mechanism. This requires some interpretation to produce a general picture of arcing, and a "complete" model of arcing becomes a series of calculations of specific mechanisms together with a general picture and set of assumptions that tie the mechanisms together. On the other hand, there seems to be sufficient knowledge of the individual mechanisms involved to produce useful predictions and explanations of experimental data.